\documentstyle[12pt]{article}

\newcommand{\be}{\begin{equation}}
\newcommand{\ee}{\end{equation}}
\newcommand{\ba}{\begin{eqnarray}}
\newcommand{\ea}{\end{eqnarray}}
\newcommand{\baa}{\begin{eqnarray*}}
\newcommand{\eaa}{\end{eqnarray*}}
\newcommand{\lab}[1]{\label{#1}}

\newcommand{\pad}{\partial}
\newcommand{\dis}{\displaystyle}
\newcommand{\non}{\nonumber}
\newcommand{\tri}{\triangle}
\newcommand{\app}{\approx}
\newcommand{\bph}{\bar{\phi}}
\newcommand{\bps}{\bar{\psi}}
\begin{document}
{\pagestyle{empty}
\rightline{SLAC-PUB-5963}
\rightline{NWU-8/92}
\rightline{October 1992}
\rightline{T/AS~~~~~~~~~~}
\vskip 1cm
{\renewcommand{\thefootnote}{\fnsymbol{footnote}}
\centerline{\Large \bf Generalized Einstein Theory on Solar
and Galactic Scales \footnote[1]{Work supported, in part,
by the Department of Energy, contract DE-AC03-76SF00515.}}}
\vskip 1cm

\centerline{Masakatsu Kenmoku \footnote{E-mail address:
kenmoku@jpnyitp.bitnet} }
\centerline{{\it Department of Physics} }
\centerline{{\it Nara Women's University, Nara 630, Japan} }

\vskip 0.1in

\centerline{Yuko Okamoto \footnote{E-mail address: yuko@slacvm ;
on leave of absence from Department of Physics, Nara Women's
University,Nara 630, Japan.}}
\centerline {{\it Stanford Linear Accelerator Center}}
\centerline {{\it Stanford University,
Stanford, CA 94309, USA}}

\vskip 0.1in

\centerline{Kazuyasu Shigemoto \footnote{E-mail address:
shigemot@jpnyitp.bitnet}}
\centerline {{\it Department of Physics}}
\centerline {{\it Tezukayama University, Nara 631, Japan }}

\vskip 1cm

\centerline{\bf Abstract}
\vskip 0.2in

We study a generalized Einstein theory with the following two
criteria:{\it i}) on the solar scale, it must be consistent with
the classical tests of general relativity, {\it ii}) on the
galactic scale, the gravitational potential is a sum of
Newtonian and Yukawa potentials so that it may explain the flat
rotation curves of spiral galaxies.
Under these criteria, we find that such a generalized Einstein
action must include at least one scalar field and one vector
field as well as the quadratic term of the scalar curvature.

\vskip 0.4cm
\centerline{Submitted to {\it Physical Review D}}
\noindent
PACS number(s): 04.20.-q, 98.80.Dr, 98.60.Eg
\hfil
\vfill
\newpage}
%%%%%%%%%%%%%%%%%%%%%%% Section 1 %%%%%%%%%%%%%%%%%%%%%%%%

\vspace{1.5cm}

\noindent
{\large \bf 1.\ \ Introduction}

%\section{Introduction}

\vspace{0.5cm}

Recent astrophysical observations of distant galaxy
distributions \cite{Broad},\cite{Geller} and the cosmic
microwave background \cite{COBE} have revealed a quite
astonishing picture of the universe.
{}From the survey of relatively near galaxies, the void structure
(the Great Wall) was discovered,\cite{Geller} and from the
pencil-beam survey of galaxies, the quasi-periodic distributions
(of period about $130Mpc$) were inferred.\cite{Broad}  The
data from COBE, on the other hand, have revealed extremely
isotropic and homogeneous distribution of the $2.7\ K$ cosmic
microwave background with fluctuations of order
$10^{-5}$.\cite{COBE} In general, it is very difficult to
explain how these anistopic and inhomogeneous large-scale
structures of the universe have developed
from such an isotropic and homogeneous distribution of matter
in the early stage of the universe.  The
standard solution to this difficulty totally relies on
the existence of dark matter which
accounts for more than 90\% of matter in the universe.
The evidence for dark matter was first claimed in order to
explain the flat rotation curves of spiral galaxies.
Since there is no established direct observation
of dark matter, however,
there are many attempts to explain the rotation curves
without dark matter by modifying
the Newtonian force \cite{Sanders} or by modifying the Newton's
second force law.\cite{Milgrom},\cite{Kent}
Other people have tried to derive such modified Newtonian
force laws from the framework of general relativity.
\cite{Brans}\cite{Fujii}\cite{Mannheim}\cite{Ours} \\

In a previous paper,\cite{Ours} we attempted to explain not
only the flat rotation curves of spiral galaxies but also
the large-scale structure of the universe, starting
from a simple model with the addition of a
quadratic scalar curvature term to the
Einstein action.
Our generalized action could qualitatively explain the
flatness of the rotation curves and the nearly periodic
galaxy distributions.
However, it turned out that our
theory does not imply the unity of the coefficient $\gamma$ of
the Robertson expansion \cite{Weinberg} on the solar
scale.\cite{Ours}
This constraint ($\gamma = 1$) from the classical tests of
general relativity such as the observation of the radar echo
delay is quite stringent, and it is very difficult to realize
this value in the generalized Einstein action.\\

In this paper, we construct the generalized Einstein
action under the two criteria:
{\it i}) it must give $\gamma=1$ in the post Newtonian
approximation,
{\it ii}) the gravitational potential is a sum of
Newtonian and Yukawa potentials.  The second criterion
is imposed, since it is the empirical gravitational potential of
Sanders \cite{Sanders} that can quite successfully
explain the flat rotation curves of spiral galaxies.
We then show that the minimum ingredient of the theory
that satisfies the above criteria
is the $R^2$ term, a scalar field, and a vector field
in addition to the Einstein action.

%%%%%%%%%%%%%%%%%%%%%%%%% section 2 %%%%%%%%%%%%%%%%%%%%%%%

\vspace{1.5cm}

\noindent
{\large \bf 2. Generalized Einstein Action and Its Post
Newtonian Approximation }

\vspace{0.5cm}

The generalized Einstein action which contains quadratic
terms of the scalar curvature, $R^2$, and Ricci tensor,
$R_{\mu \nu}R^{\mu \nu}$, was
introduced to regulate the ultraviolet
divergences of the Einstein theory.\cite{Utiyama}  It was
applied to cosmology to obtain the bounce universe to avoid
the singularity at the
creation of the universe.\cite{Nariai}
The structure and the properties of the theory
were further elaborated in subsequent works. \cite{Higher} \\

In this section, we consider a further generalization of the
theory by adding scalar and vector fields
in addition to $R^2$ and $R_{\mu \nu}R^{\mu \nu}$ terms and
study its post Newtonian approximation.  We will investigate
such a theory with two requirements:
{\it i}) on the solar scale, it must conform with the
classical tests of general relativity, {\it ii}) on the galactic
scale, the gravitational potential is modified to give a sum of
Newtonian and Yukawa potentials in order
to explain the rotational velocity
curves of spiral galaxies. \\

We consider the following generalized action with the scalar
curvature $R$, the Ricci tensor $R_{\mu\nu}$,
a scalar field $\sigma$, and a vector field $A_{\mu}$:
\ba
I&~=\dis{ \int} d^4x \sqrt{-g} \Bigl\{ -\dis{ \frac{1}{16 \pi G}}
(R + c_1 R^2 + c_2 R^{\mu \nu} R_{\mu \nu} )  \non \\
& -\dis{\frac{1}{2}} \pad_{\mu}\sigma \pad^{\mu}\sigma
- \dis{ \frac{\mu^2}{2}} \sigma^2
+g_1 \sigma R                             \lab{e1} \\
&  -\dis{\frac{1}{2}} D_{\mu}A_{\nu} D^{\mu}A^{\nu}
- \dis{\frac{m^2}{2}} A^{\mu}A_{\mu}
+g_2 D_{\mu}A^{\mu} R + L^{matter}  \Bigr\}~, \non \\
\non
\ea
where $G$ is the gravitational constant,
$\mu$ and $m$ are the masses of the scalar and vector particles,
$D_{\mu}$ is the covariant derivative, and $L^{matter}$
is the matter Lagrangian. The coefficients $c_1, c_2$, and $G$
have the dimension of $(mass)^{-2}$, while the coefficient
$g_1$ has the dimension of mass and $g_2$ is dimensionless. \\

In order to calculate the coefficient $\gamma$
in the Robertson expansion,\cite{Weinberg}
we introduce the weak fields $\phi$ and $\psi$ defined by
\be
g_{0 0}=-1-2\phi, \quad g_{ij}=\delta_{ij}(1+2\psi)~. \lab{e2}
\ee
In the post Newtonian approximation, we must take into account
up to the quadratic term of the weak fields and source in the
action, and the
necessary formulae to the first order in the weak fields are
\ba
& \sqrt{-g} \app 1+\phi+3\psi~,\quad R_{00} \app -\tri\phi~,
\quad \non \\
&              \non   \\
& R_{ij} \app \pad_i\pad_j (\phi+\psi)+\delta_{ij}\tri\psi~,
\quad
R \app 2\tri(\phi+2\psi)~, \lab{e3} \\
\non
\ea
and the necessary formulae up to the second order in the weak
fields and source in the action are
\ba
& \sqrt{-g}R \app 2\tri(\phi+2\psi)
+4\phi\tri\psi+2\psi\tri\psi~, \non  \\
&                 \non  \\
& \sqrt{-g}R^{\mu \nu}R_{\mu \nu} \app 2(\tri\phi)^2
+4\tri\phi\tri\psi +6(\tri\psi)^2~, \lab{e4} \\
&                 \non  \\
& \sqrt{-g}R^2 \app 4\{\tri(\phi+2\psi)\}^2~, \quad
\sqrt{-g}L^{matter} \app -\rho\phi~, \non \\
\non
\ea
where we have suppressed
the total derivative terms.\cite{Veltman}
We substitute the weak field expressions Eqs.~(\ref{e3})
and (\ref{e4}) into Eq.~(\ref{e1}), and retain up to the
quadratic terms of the weak fields $\phi,\psi,\sigma,
A_{\mu}$ and source $\rho$ to obtain
\ba
I \app & \dis{\int} d^4x  \Bigl[ -\dis{\frac{1}{16\pi G}}
\bigl\{2\tri(\phi+2\psi)+4\phi\tri\psi+2\psi\tri\psi
+4c_1(\tri(\phi+2\psi))^2      \non \\
&+4c_2(\dis{\frac12}(\tri\phi)^2+\tri\phi\tri\psi
+\dis{\frac32}(\tri\psi)^2 \bigr\}       \non \\
&+\dis{\frac12}\sigma\tri\sigma-\dis{\frac{\mu^2}2}\sigma^2
+\dis{\frac12}A_i\tri A_i-\dis{\frac{m^2}2}A_i^2
-\dis{\frac12}A_0\tri A_0+\dis{\frac{m^2}2}A_0^2  \lab{e5} \\
&                   \non       \\
&+2g_1\sigma\tri(\phi+2\psi)
+2g_2\pad_iA_i\tri(\phi+2\psi)-\rho\phi \Bigr]~. \non
\ea
Since the field $A_0$ decouples from the other fields and
source, we do not consider this field hereafter.
It is convenient to introduce the following
variables $\bph$ and $\bps$:
\be
\bph=\phi+2\psi~, \quad \bps=\psi-\phi~, \lab{e6}
\ee
which in turn gives
\be
\phi=\dis{\frac{\bph-2\bps}{3}}~, \quad \psi=
\dis{\frac{\bph+\bps}{3}}~.\lab{e7}
\ee
Substituting Eq.~(\ref{e7}) into Eq.~(\ref{e5}), we can
rewrite the action as
\ba
I \app & \dis{\int} d^4x  \Bigl[ -\dis{\frac{1}{16\pi G}} \bigl\{
2\tri\bph+\dis{\frac23}\bph\tri\bph-\dis{\frac23}\bps\tri\bps
+4c_1(\tri\bph)^2 +\dis{\frac{4c_2}{3}}((\tri\bph)^2
+\dis{\frac12}(\tri\bps)^2)\bigr\} \non \\
&+\dis{\frac12}\sigma\tri\sigma-\dis{\frac{\mu^2}2}\sigma^2
+\dis{\frac12}A_i\tri A_i-\dis{\frac{m^2}2}A_i^2
+2g_1\sigma\tri\bph+2g_2\pad_iA_i\tri\bph
-\dis{\frac{\rho\bph}3}+\dis{\frac{2\rho\bps}3} \Bigr]~.
\lab{e8} \\ \non
\ea
Taking the variation with respect to the weak fields $\bph,
\bps, A_i $ and $\sigma$, we obtain the equation of motion of
the weak fields in the post Newtonian approximation:
\ba
&\tri\left\{1+(6c_1+2c_2)\tri\right\}\bph-24\pi
G(g_1\tri\sigma+g_2\tri\pad_i A_i )=-4\pi G \rho~,  \lab{e9} \\
&                   \non      \\
& \tri(1-c_2\tri)\bps=-8\pi G \rho~,   \lab{e10} \\
&                   \non      \\
& (\tri-m^2) A_i-2g_2\tri\pad_i\bph=0~,  \lab{e11} \\
&                   \non      \\
& (\tri-\mu^2) \sigma+2g_1\tri\bph=0~. \lab{e12} \\
\non
\ea
{}From Eqs.~(\ref{e11}) and (\ref{e12}), we have
\be \pad_i A_i=\dis{\frac{2g_2\tri^2\bph}{\tri-m^2}}~, \quad
  \sigma=-\dis{\frac{2g_1\tri\bph}{\tri-\mu^2}}~.   \lab{e13}
\ee
Substituting this expression into Eqs.(\ref{e9}) and
(\ref{e10}), we then obtain the following:
\ba
& \tri\left\{1+(6c_1+2c_2)\tri-48\pi
G(-\dis{\frac{g_1^2\tri}{\tri-\mu^2}}+
\dis{\frac{g_2^2\tri^2}{\tri-m^2}})\right\}\bph=-4\pi G\rho~,
\lab{e14} \\
&                    \non       \\
& \tri(1-c_2\tri) \bps=-8\pi G \rho~. \lab{e15} \\
\non
\ea
In the next section, we will consider the possibility that
the coefficient of the Robertson expansion $\gamma$ becomes $1$.
In order to obtain this result, it is necessary that both
$\bph$ and $\bps$ behave $\sim 1/r$ in the limit
$r \rightarrow 0$. ( We remark from Eq.~(\ref{e7}):
if $\bph$ and $\bps$ behave like
$\bph\sim{\rm const.}$ and $\bps\sim 1/r$ in the limit
$r \rightarrow 0$,
we obtain $\gamma=1/2$, while if $\bph$ and $\bps$ behave
like $\bph\sim 1/r$ and $\bps\sim {\rm const.}$ in the
limit $r \rightarrow 0$,
we obtain $\gamma=-1$.)
In the region $r \rightarrow 0$, the mass term is
negligible, and the necessary condition to have $\gamma=1$
is that the highest derivative terms ($\propto \tri^2$)
on the left-hand sides of Eqs.~(\ref{e14}) and (\ref{e15})
must vanish (for $\mu^2=m^2=0$). This condition reads
\be
6c_1+2c_2-48\pi G g_2^2=0~, \qquad c_2=0~.  \lab{e16}
\ee
Near the origin we then have
\ba
 (1-6 c_1 m^2+48 \pi G g_1^2)\tri\bph &\app -4\pi G \rho~,
\lab{e17} \\
&                \non   \\
  \tri\bps & \app -8\pi G \rho~. \lab{e18} \\
\non
\ea
We assume that the density takes the point-like
distribution of the form
$$ \rho( \vec{r}) = M \delta(\vec{r})~. $$
Using the formula $ 4\pi G \rho /\tri =- G M/r $,
we then obtain
\ba
&\bph \app -\dis{\frac{4\pi G \rho}{k \tri}}=
\dis{\frac{G M}{k r}}~, \lab{e19} \\
&\bps \app -\dis{\frac{8\pi G \rho}{\tri}}=
\dis{\frac{2G M}{r}}~, \lab{e20} \\
\non
\ea
where
\be
\quad k=1-6c_1 m^2+48 \pi G g_1^2~. \lab{e21} \\
\ee
Therefore we obtain the gravitational potential of the form
\ba
& \phi\app\dis{\frac{\bph-2\bps}{3}}
=-\dis{\frac{G M}{r}}\dis{(\frac{4k-1}{3k})}
\equiv -\dis{\frac{\bar{G} M}{r}}~, \lab{e22} \\
& \psi\app\dis{\frac{\bph+\bps}{3}}
=\dis{\frac{G M}{r}}\dis{(\frac{2k+1}{3k})}
\equiv \dis{\frac{\bar{G} M}{r} \gamma}~, \lab{e23} \\
\non
\ea
where
\be
\quad \bar{G}=\dis{\frac{4k-1}{3k}}G~. \lab{e24} \\
\ee
This leads to the following formula for $\gamma$~:
\be
\gamma=\dis{\frac{2k+1}{4k-1}}~. \lab{e25} \\
\ee
Here, by taking Morikawa model \cite{Morikawa} as an
example, we demonstrate how it is difficult to satisfy the
stringent condition $\gamma=1$ in a modified Einstein
theory in general. \\

Morikawa model is a Brans-Dicke type theory of
the form \cite{Morikawa}
\ba
I^{Morikawa}&=\dis{ \int} d^4x \sqrt{-g} \Bigl\{
-\dis{ \frac{1}{16 \pi G}}(R -2\Lambda)  \non \\
&-\dis{\frac{1}{2}} \pad_{\mu}\varphi \pad^{\mu}\varphi
- \dis{ \frac{\mu^2}{2}} \varphi^2 +\lambda\varphi
+\dis{ \frac{\xi\varphi^2}{2}}R + L^{matter}  \Bigr\}~,
\lab{e26} \\
\non
\ea
where we have added the cosmological term $\Lambda$
and tadpole term $\lambda\varphi$ to the original Morikawa
action, since we consider the case that the
scalar field $\varphi$ takes an expectation value $v$.
We write $\varphi=v+\sigma$ and consider this $\sigma$ field
as the weak field. We then obtain the weak field
approximation of the form
\ba
I^{Morikawa} & \app \dis{ \int} d^4x \sqrt{-g}
\Bigl\{ -\dis{ \frac{R}{16 \pi G}}(1-8\pi G \xi v^2)  \non \\
&-\dis{\frac{1}{2}} \pad_{\mu}\sigma \pad^{\mu}\sigma
- \dis{ \frac{\mu^2}{2}} \sigma^2 + \xi v \sigma R + L^{matter}
                                  \Bigr\}~.       \lab{e27} \\
\non
\ea
Here, we have tuned $\Lambda$ and $\lambda$ in such a way that
$$
\dis{( \frac{\Lambda}{8 \pi G}-\frac{\mu^2v^2}{2}+\lambda v)}=0~,
\quad \lambda-\mu^2 v=0~,
$$
for a given $v$.
If we denote $ 1/{\tilde{G}}=(1-8\pi G \xi v^2)/G $, then
according to our formula Eq.~(\ref{e21})
(with the replacements $m \rightarrow 0,
G \rightarrow \tilde{G}$, and $g_1 \rightarrow \xi v$),
we have the expression
$k=1+48\pi \tilde{G} \xi^2 v^2$, which gives
\be
\gamma=\dis{\frac{1-8\pi G (\xi-4 \xi^2) v^2}
{1-8\pi G (\xi-8 \xi^2) v^2}}~.     \lab{e28}
\ee
In Morikawa's case, \cite{Morikawa}
the scale factor of the universe oscillate with time,
where the period is converted to
the scale of $130 Mpc$ through the velocity of light,
and it is unclear whether
$\varphi$ here takes an expectation value on
the solar scale.  We obtain
$\gamma=0.88$ by using Morikawa's unit $1/G=4\pi/3 $ and
his values $\xi=10$ and $v=0.008$, where
the value of $v$ is assumed to be
of the order of the initial value of $\varphi$~.
(Of course, if we set $v=0$
in Morikawa model, we have $\gamma=1$.) \\

Through this example, we understand that the construction of the
generalized Einstein action with $\gamma=1$ is quite non-trivial.

%%%%%%%%%%%%%%%%%%%%%%%%%%% section 3 %%%%%%%%%%%%%%%%%%%%%%%%%%

\vspace{1.5cm}

\noindent
{\large \bf 3. Generalized Einstein Action on Solar and
Galactic Scales}

\vspace{0.5cm}
In this section, we show that
the generalized Einstein theory, in which
the gravitational potential is a sum of
Newtonian and Yukawa potentials,
must include at least
scalar and vector fields in addition to
the quadratic term of the scalar curvature
in order that the condition $\gamma=1$ is satisfied. \\

The condition $\gamma=1$ implies $k=1$
in Eq.~(\ref{e25}), which in turn implies
$ c_1 m^2=8 \pi G g_1^2$ (see Eq.~(\ref{e21})).
Taking into account Eq.~(\ref{e16}) also, the necessary
condition to have $\gamma=1$ is summarized as follows:
\ba
c_1&=&8\pi G g_2^2,\quad c_2=0~,\lab{e29}  \\
c_1 m^2&=&8 \pi G g_1^2~.   \lab{e30}    \\
\non
\ea

This condition is classified into the following possibilities:
$$
\begin{array}{cccc}
    {\it I})    & g_2\ne 0~,&\quad    & \quad \\
    \quad     &{\it a})  & g_1= 0~, & (c_1\ne 0,\ m=0),\\
    \quad    &{\it b}) & g_1\ne 0~, & (c_1\ne 0,\ m\ne0),\\
     {\it II})   & g_2= 0~, & \quad  & (c_1=0,\ g_1=0).\
\end{array}
$$
We first consider case {\it Ia}), where the scalar decouples
from other fields.
In this case Eqs.~(\ref{e14}) and (\ref{e15}) become
\be
\tri\bph=-4\pi G \rho, \quad \tri\bps=-8\pi G \rho~,\lab{e31}
\ee
which in turn gives
\be
\bph=-\dis{\frac{4\pi G \rho}{\tri}=\frac{GM}{r}}~, \quad
\bps=-\dis{\frac{8\pi G \rho}{\tri}=\frac{2GM}{r}}~. \lab{e32}
\ee
Hence, we obtain the same result as the Newtonian
approximation of the ordinary Einstein theory:
\be
\dis{\phi=-\frac{GM}{r}}~,\quad \dis{\psi=\frac{GM}{r}}~,
\lab{e33}
\ee
in which the desired Yukawa term is absent. \\

In case {\it II}), the
vector field decouples and there is no higher
derivative term such as $R^2$ and $R_{\mu \nu}R^{\mu \nu}$.
In this case, Eq.~(\ref{e14}) takes the same form as
Eq.~(\ref{e31}), and we obtain the result of the Newtonian
approximation again. \\

The final case {\it Ib}) turns out the one that satisfies
our criteria.
In this case, from the relation $c_1m^2=8\pi G g_1^2$,
Eq.~(\ref{e14}) becomes
\ba
& \dis{\frac{\tri}{(\tri-m^2)(\tri-\mu^2)}}\Bigl\{\tri^2
-\tri\left(m^2+\mu^2+48\pi G g_1^2(m^2-\mu^2)\right)
\non \\
&+m^2\mu^2 \Bigr\}\bph=-4\pi G \rho~. \lab{e34}
\ea
We then define $\alpha$\ and $\beta$\ by
\ba
\alpha+\beta &=& m^2+\mu^2+48\pi G g_1^2(m^2-\mu^2)~,\non \\
\alpha \beta &=& m^2\mu^2~,\lab{e35}
\ea
and assume $\alpha>\beta$.  We can also
define constants $k_1, k_2$\ and $k_3$ by
\ba
\bph &=& -\dis{\frac{4\pi G(\tri-m^2)(\tri-\mu^2)}
{\tri(\tri-\alpha)(\tri-\beta)}}\rho~, \non \\
&=&-4\pi G \Bigl\{ \dis
\dis{\frac{k_1}{\tri}}+\dis{\frac{k_2}{\tri-\alpha}}
+\dis{\frac{k_3}{\tri-\beta}} \Bigr\} \rho~,
\lab{e36} \\
\non
\ea
which gives the relations among
$k_1, k_2 $\ and $k_3$ as follows:
\ba
&& k_1+k_2+k_3=1~, \non \\
&& k_1(\alpha+\beta)+k_2 \beta+k_3 \alpha=m^2+\mu^2~,\lab{e37} \\
&& k_1 \alpha \beta=m^2 \mu^2~.      \non
\ea
{}From Eq.~(\ref{e35}) we obtain
$k_1=1$, so that $k_2$ and $k_3$ are related by $k_3=-k_2$.
Using this relation and substituting Eq.~(\ref{e35}) into
Eq.~(\ref{e37}), we obtain
\be
k_2=-k_3=\dis{\frac{48\pi G g_1^2(m^2-\mu^2)}{\alpha-\beta}}~.
\lab{e38}
\ee
The modified gravitational potential in this case becomes
\ba
\bph &=&-4\pi G \dis{\left( \frac{1}{\tri}+\frac{k_2}{\tri-\alpha}
-\frac{k_2}{\tri-\beta} \right)}\rho~,\non \\
&=& \dis{\frac{GM}{r}}\left(1+k_2e^{-\sqrt{\alpha} r}
-k_2e^{-\sqrt{\beta} r}\right)~, \lab{e39} \\
\bps &=& \dis{\frac{2GM}{r}}~. \non
\ea
Therefore we have the gravitational potential of
the desired form:
\be
\phi=\dis{\frac{\bph-2\bps}3}=-\dis{\frac{GM}r}
\left( 1-\dis{\frac{k_2}3} e^{-\sqrt{\alpha} r}
   +\dis{\frac{k_2}3} e^{-\sqrt{\beta} r} \right)~.
\lab{e40}
\ee

We now examine the condition for this gravitational potential
to explain the flat rotation curves of spiral galaxies.
We know that the gravitational potential in Sanders'
form \cite{Sanders}
\be
\phi=-\dis{\frac{GM}r}\left(\frac{1+\alpha_S~e^{-ar}}
{1+\alpha_S}\right) \lab{e41}
\ee
can account for the rotation curves in a satisfactory way when
$\alpha_S=-0.9$.  For our potential Eq.~(\ref{e40}) to have a
similar form on the galactic scale, it is necessary to
assume (for $\alpha > \beta > 0$)
\ba
\sqrt{\alpha}&=&{\cal O}\left(\frac{1}{r_0}\right)~, \lab{e42} \\
\alpha &\gg& \beta~>~0~, \lab{e43}
\ea
where $r_0$ is a distance on galactic scale
($\sim$ a few 10$kpc$).
Note that these relations imply
\be
\exp{(-\sqrt{\beta} r_0)}\app 1~. \lab{e44}
\ee
For the rotation curves to be flatter than the Newtonian results,
we need $\alpha_S < 0$ in Eq.~(\ref{e41}).  In our potential
Eq.~(\ref{e40}) this requirement together with Eqs.~(\ref{e43})
and (\ref{e44}) implies
\be
k_2 > 0~. \lab{e45}
\ee
By Eq.~(\ref{e38}) this is equivalent to
\be
m^2 > \mu^2~. \lab{e46}
\ee
Hence, the question is whether we can satisfy the above
conditions Eqs.~(\ref{e42}), (\ref{e43}), and (\ref{e45})
(or (\ref{e46})) in a consistent way.  From Eqs.~(\ref{e35})
and (\ref{e46}), one obvious solution is
\ba
\alpha&=&{\cal O}\left(m^2\right)~, \non \\
\beta&=&{\cal O}\left(\mu^2\right)~, \lab{e47} \\
m^2&=&{\cal O}\left(\frac{1}{r_0^2}\right) \gg \mu^2~, \non
\ea
where $r_0$ is the galactic scale.
In this case, from Eq.~(\ref{e35}) $\alpha$ can be written as
\be
\alpha \app m^2(1+48\pi G g_1^2)~. \lab{e48}
\ee
By defining
$x=48\pi G g_1^2(>0)$ and neglecting $\beta$ and $\mu^2$ in
Eq.~(\ref{e38}), we have
\be
k_2=\frac{x}{1+x}~, \lab{e49}
\ee
so that $k_2$ can change in the range
\be
0 < k_2 < 1~. \lab{e50}
\ee
The coefficient $\alpha_S$ of Eq.~(\ref{e41}) in our theory
is then written as
\be
\alpha_S = - \frac{k_2}{3+k_2} = - \frac{x}{3+4x}~. \lab{e51}
\ee
Note that our $\alpha_S$ takes a value in the range
\be
-0.25<\alpha_S<0~, \lab{e52}
\ee
while Sanders takes $\alpha_S=-0.9$ to fit the rotation curves. \\

In this way, starting from the generalized
Einstein action with additional
scalar and vector fields, we can tune the parameters so that
we not only have $\gamma=1$
on the solar scale but also obtain
the empirical Sanders type
gravitational potential on the galactic scale.

%%%%%%%%%%%%%%%%%%%%%%%%%%% section 4 %%%%%%%%%%%%%%%%%%%%%%

\vspace{1.5cm}

\noindent

{\large \bf 4. Summary and Discussion}

\vspace{0.5cm}

In this paper, we have examined the generalized Einstein theory
which contains higher derivative terms, $R^2$ and
$R^{\mu \nu}R_{\mu \nu}$,
and satisfies the criteria:
{\it i}) on the solar scale, it must be consistent with the
classical tests of general relativity, {\it ii}) on the galactic
scale, the gravitational potential is a sum of Newtonian and
Yukawa potentials so that it may
explain the flat rotation curves of spiral galaxies.
We have shown that it is non-trivial to satisfy the above
criteria and that at least additional
scalar and vector fields are required for a consistent theory. \\

We have tuned the parameters of the theory so that the
coefficient $\alpha_S$ of the Yukawa term ( in Sanders'
gravitational potential) is negative, which is necessary to
explain the flatness of rotation curves of spiral galaxies.
It will be interesting to see how well our theory can fit the
rotation curves quantitatively.
Numerical calculations in this direction are now in  progress. \\

In our generalized Einstein action, even after the tuning of
parameters, there are still two
parameters, the vector mass $m$ and the scalar $\mu$, to set the
scale of interest.
We have chosen $1/m\sim $ galaxy scale, and it may be
possible to explain the \lq \lq periodic" large-scale structure
of the universe by choosing $1/\mu \sim 130Mpc$.
Numerical calculations of distant galaxy distributions are also
in progress. \\

Therefore, by taking the
values of $1/m$ and $1/\mu$ as above, our generalized Einstein
theory may be consistent with observations over three different
distance scales: the solar, galactic and
beyond galactic scales of the universe.

\vspace{1.5cm}
\noindent
{\large \bf Acknowledgements}:
  One of us (Y.O.) is grateful to the members of Stanford Linear
Accelerator Center for hospitality.
\vspace{1.0cm}
\newpage

%%%%%%%%%%%%%%%%%%%%%%%%% references %%%%%%%%%%%%%%%%%%%


\noindent
\begin{thebibliography}{[00]}

\bibitem{Broad} T.J. Broadhurst, R.S. Ellis, D.C. Koo and
         A.S. Szalay, Nature, {\bf 343},726~(1990).
\bibitem{Geller} M.J. Geller and J.P. Huchra, Nature,
         {\bf 246}, 897~(1989).
\bibitem{COBE} G.F. Smoot $et.al.$, Astrophys. J.
         {\bf 396}, L1~(1992).
\bibitem{Sanders} R.H. Sanders, Astron. Astrophys. J.
         {\bf 136}, L21~(1984)
\bibitem{Milgrom} M. Milgrom, Astrophys. J. {\bf 270}, 365~(1983).
\bibitem{Kent} S.M. Kent, Astron. J. {\bf 93}, 816~(1987).
\bibitem{Brans} C. Brans and R.H. Dicke, Phys. Rev.
         {\bf 124}, 925~(1961).
\bibitem{Fujii} Y. Fujii, Phys. Rev. D{\bf 9}, 874~(1974).
\bibitem{Mannheim} P.D. Mannheim and D. Kazanas, Astrophys. J.
         {\bf 342}, 635~(1989).
\bibitem{Ours} M. Kenmoku, E. Kitajima, Y. Okamoto and
         K. Shigemoto, Int. J. Mod. Phys. D, to be published.
\bibitem{Weinberg} S. Weinberg, \lq\lq Gravitation and
         Cosmology"~(John Wiley and Sons, New York, 1972).
\bibitem{Utiyama} R. Utiyama and B.S. de Witt, J. Math. Phys.
         {\bf 3}, 608~(1962).
\bibitem{Nariai} H. Nariai and K. Tomita, Progr. Theor. Phys.
         {\bf 46}, 776~(1971)
\bibitem{Higher} K. Stella, Gen. Rel.Grav. {\bf 9}, 353~(1978);
         B. Whitt, Phys.Lett. {\bf 145B}, 176~(1984);
          R. Utiyama, Progr. Theor. Phys.{\bf 72}, 83~(1984).
\bibitem{Veltman} M.J.G. Veltman, in \lq\lq Methods in Field
          Theory"~(Les Houches, 1975) eds. R. Balian and
          J. Zinn-Justin~(North-Holland, Amsterdam, 1976).
\bibitem{Morikawa} M. Morikawa, Astrophys. J. {\bf 369},
          20~(1991).
\end{thebibliography}
\end{document}